\documentclass{ws-p8-50x6-00}

\begin{document}

\title{Monte Carlo Hamiltonian: Generalization to Quantum Field Theory}

\author{Xiang-Qian Luo}
\address{Department of Physics, Zhongshan University
Guangzhou 510275, China}

\author{Hamza Jirari, Helmut Kr\"oger}
\address{D\'epartement de Physique, Universit\'e Laval, Qu\'ebec, 
Qu\'ebec G1K 7P4, Canada}

\author{Kevin J.M. Moriarty}
\address{Department of Mathematics, Statistics and Computer Science,
  Dalhousie University, Halifax, Nova Scotia B3H 3J5, Canada}


\maketitle

\abstracts{Monte Carlo techniques with importance sampling 
have been extensively applied to lattice gauge theory in the Lagrangian formulation.
Unfortunately, it is extremely difficult to compute 
the excited states using the conventional Monte Carlo algorithm.
Our recently developed approach: the Monte Carlo Hamiltonian
method, has been designed to overcome the difficulties of the
conventional approach. In this paper, 
we extend the method to many body systems and quantum field theory.
The Klein-Gordon field theory is used as a testing ground.}

\section{Introduction}

Quantum theory is one of the most important achievements 
in modern science, but 
a lot of non-perturbative aspects are still poorly understood.
There are two standard formulations in quantum theory:
Hamiltonian and Lagrangian.
A comparison of the conventional approaches is given in Tab.\ref{tab1}.
 
The Lagrangian formulation is very suitable 
for applying the Monte Carlo (MC) method to
systems with many degrees of freedom,
and in the last two decades, it has been widely applied 
to lattice gauge theory\cite{Creutz,Rothe,Montvay}.
In the standard Lagrangian MC method, however,
it is extremely difficult to compute the spectrum 
and wave function beyond the ground state.
On the other hand, the standard Hamiltonian formulation is capable of
doing it. 
If fact, some efforts have been made to solve the lattice QCD Schr\"odinger equation\cite{Guo:1997sc}.
But due to the complexity of non-Abelian gauge theory and high degrees of freedom,
only the ground state\cite{Chen:1995a}
and the lowest lying glueball states\cite{Chen:1995b,Luo:1996ha,Luo:1997sa} have been investigated. 
There have also been some attempts to perform numerical simulations
in the Hamiltonian formulation, e.g. 
the quantum MC technique\cite{Honma}, 
MC method using a guided random walk\cite{Chin,Barnes}, and 
the MC Green's function method\cite{Beccaria,Hamer}. 
For an review of MC methods applied 
to solve the quantum many-body problem 
see Ref.\cite{Linden}.

A natural question is whether in Lagrangian MC simulations 
one can construct an effective Hamiltonian?
If yes,  the excited states can also be computed.
We have recently proposed a new approach\cite{mch} 
(we called it Monte Carlo Hamiltonian method) to investigate this problem.
A lot of models\cite{mch,physica,lat99_1,Jiang,lat99_2,Huang} 
in quantum mechanics (QM) have been used to test the method, as briefly reviewed
in Sect. \ref{review}.
In this paper, 
we will extend the method to many body systems and quantum field theory (QFT).

\begin{table}
\caption{Comparison of the conventional methods in the standard formulations.}
\begin{center}
\begin{tabular}{|c|c|c|}
\hline
{\bf  Formulation} & {\bf Hamiltonian}  &  {\bf Lagrangian} \\
\hline
Approach & Schr\"odinger Eq.                    & Path Integral\\
         & $H \vert E_n \rangle = E_n \vert E_n \rangle$
                                                &   $\langle O \rangle ={\int [d \varphi] O[\varphi]
                                                    \exp(- S[\varphi]/ \hbar)  \over
                                             \int [d \varphi] \exp(- S[\varphi]/ \hbar)}$ \\
\hline
Algorithm & Series expansion,               & MC simulation      \\
          & variational,                    &                    \\
          & Runge-Kutta  ...                &                    \\
\hline
Advantage  & Both the ground state,       & It generates the most    \\
           & and the excited states       & important configs.  \\ 
           & can be computed.             & for the measurements.   \\
\hline
Disadvantage & Analytical methods         & It is difficult to study \\
             & are too tedious for        & the excited states, and  \\
             & many body systems;         & finite density QCD.      \\
             & Runge-Kutta works          &                          \\
             & only in 1-D.               &                          \\
\hline
            
\end{tabular}
\end{center}
\label{tab1}
\end{table}


\section{Algorithm}
\label{review}

\subsection{Effective Hamiltonian}
\label{EH}

The basic idea of the MC Hamiltonian method for one body quantum mechanics
has been described in details in Ref.\cite{mch}.
The (imaginary time) transition amplitude between 
an initial state at position  
$x_i$, and time $t_i$,  
and final state at $x_f$ and $t_f$ is related 
to the Hamiltonian $H$ by
\begin{eqnarray}
M_{fi} &=& \langle x_{f},t_{f}  \vert  x_{i},t_{i} \rangle 
= \langle x_{f}  \vert  e^{-H(t_{f}-t_{i})/\hbar}  \vert  x_{i} \rangle
\nonumber \\
&=& \sum_{n=1}^{\infty} \langle x_{f}  \vert  E_{n}  \rangle 
e^{-E_{n} T/\hbar} \langle  E_{n}  \vert  x_{i} \rangle,
\end{eqnarray}
where $T=t_f-t_i$.
According to Feynman's path integral formulation of QM,
the transition amplitude is also related to the path integral:
\begin{eqnarray}
M_{fi} =
\int [dx] \exp(- S[x]/ \hbar) \vert_{x_i,t_i}^{x_f,t_f},
\end{eqnarray}
where $S=S_0+S_V$ is the action for a given path, 
$S_0 = \int_{t_i}^{t_f} dt ~ m \dot{x}^2 /2$,
and $S_V = \int_{t_i}^{t_f} dt ~ V(x)$.

The results above are standard ones in Lagrangian formulations.
What is new in our method\cite{mch}
is to construct an effective Hamiltonian $H_{\rm{eff}}$ 
(finite $N \times N$ matrix) through
\begin{eqnarray}
M_{fi} =
\langle x_{f}  \vert  e^{-H_{\rm{eff}} T/\hbar}  \vert  x_{i} \rangle
=\sum_{n=1}^{N} \langle  x_{f}  \vert  E^{\rm{eff}}_{n} \rangle e^{-E^{\rm{eff}}_{n} T/\hbar} 
\langle  E^{\rm{eff}}_{n}  \vert  x_{i}  \rangle .
\end{eqnarray}
The eigenvalues $E^{\rm{eff}}_{n}$ and wave function
$\vert E^{\rm{eff}}_{n}  \rangle$  can be obtained,
by diagonalizing $M$ using a unitary transformation
%
$M = U^{\dagger}DU$,
%
where $D =diag (e^{-E^{\rm{eff}}_{1}T/\hbar},..., e^{-E^{\rm{eff}}_{N}T/\hbar})$.
Once the spectrum
and wave functions are available, 
all physical information can also be obtained.

Since the theory described by $H$, which has an infinite basis in
Hilbert space, 
is now approximated by a theory
described by a finite matrix $H_{\rm{eff}}$, which has a finite basis,
the physics
of $H$ and $H_{\rm{eff}}$ might be quite different at high energy. 
Therefore we expect that we can only
reproduce the low energy physics of the system.
This is good enough for our purpose. 
In Refs.\cite{mch,physica,lat99_1,Jiang,lat99_2,Huang}, 
we investigated many 1-D, 2-D and 3-D QM 
models (Tab.\ref{tab.2}) using this MC Hamiltonian algorithm.
We computed the spectrum, wave functions and some thermodynamical observables.
The results are in very good agreement with those from analytical and/or Runge-Kutta methods.

\begin{table}[hbt]
\caption{QM systems, investigated by the MC Hamiltonian method with a regular basis.}
\begin{center}
\begin{tabular}{|c|c|c|}
\hline
System & Potential\\
\hline

QM in 1-D	& $V(x)=0$\\
                & $V(x)={1\over 2} m \omega^2 x^2$\\ 
        	& $V(x)=-V_0 \rm{sech}^2(x)$\\
        	& $V(x)={1\over 2} x^2+{1\over 4} x^4$\\
        	& $V(x)={1 \over 2} \vert x \vert$\\
                & $V(x) =\{
                       \begin{array}{cc}
                         \infty,        & x < 0 \\
                         Fx,            &  x \ge  0 
                  \end{array}$\\ 
\hline
QM in 2-D   & $V(x,y)={1\over 2} m \omega^2 x^2+{1\over 2} m \omega^2 y^2$\\
              & $V(x,y)={1\over 2} m \omega^2 x^2+{1\over 2} m \omega^2 y^2+\lambda xy$\\
\hline
QM in 3-D  & $V(x,y,z)={1\over 2} m \omega^2 x^2+{1\over 2} m \omega^2 y^2+{1\over 2} m \omega^2 z^2$\\
\hline
\end{tabular}
\end{center}
\label{tab.2}
\end{table}

\subsection{Basis in Hilbert Space}

To get the correct scale for the spectrum, the 
position state $\vert x_n \rangle$   
at initial time $t_i$ or final time $t_f$
(Bargman states or box states) 
should be properly normalized. We denote a normalized basis 
of Hilbert states as 
$\vert e_n \rangle$, $n=1, ..., N$. In position space, it can be
expressed as
\begin{eqnarray}
e_n(x) =\{
\begin{array}{cc}
     1/\sqrt{\Delta x_n},        & x \in [x_n,x_{n+1}] \\
     0,                         &  x \notin  [x_n,x_{n+1}] 
\end{array}
\end{eqnarray}
where $\Delta x_n=x_{n+1}-x_n$.

There is some arbitrariness in choosing  a basis for the
initial and final states.  
The simplest choice is a basis with $\Delta x_n ={\rm const.}$, 
which is called the ``regular basis''.
In Refs.\cite{mch,physica,lat99_1,Jiang}, the regular basis was used.
For many body systems or QFT, the regular basis will encounter problems.
For example, in a system with a 1-D chain of oscillators (see later), 
if the number of oscillators
is 30, the minimum non-trivial regular basis 
is $N=2^{30}=1073741824 \approx 10^9$, 
which is prohibitively large
for numerical calculations of the matrix elements.

Guided by the idea of importance sampling, in Refs.\cite{lat99_2,Huang}, 
we proposed to select a basis
from the Boltzmann weight proportional 
to the transition amplitude between $x'_i=0$ at $t'_i=0$ 
and $x'_f=x_n$ at some $t'_f$.
In a free particle or harmonic oscillator case, the distribution is just a Gaussian
\begin{eqnarray}
P_{basis}[x_n] ={1 \over \sqrt{2\pi} \sigma} 
\exp \left( -{x_n^2 \over 2 \sigma^2} \right),
\label{STO}
\end{eqnarray}
where $\sigma=\sqrt{\hbar T'/m}$ for the free case 
and  $\sigma=\sqrt{\hbar {\rm sinh}(\omega T')/(m \omega)}$ 
for the harmonic oscillator. 
We call such a basis the ``stochastic basis'', to be
used as states at initial time $t_i$ or final time $t_f$.  
For more complicated models, one may still use the Gaussian distribution,
but with an adjustable parameter $\sigma$.

\begin{figure}[htb]
\centerline{\epsfig{file=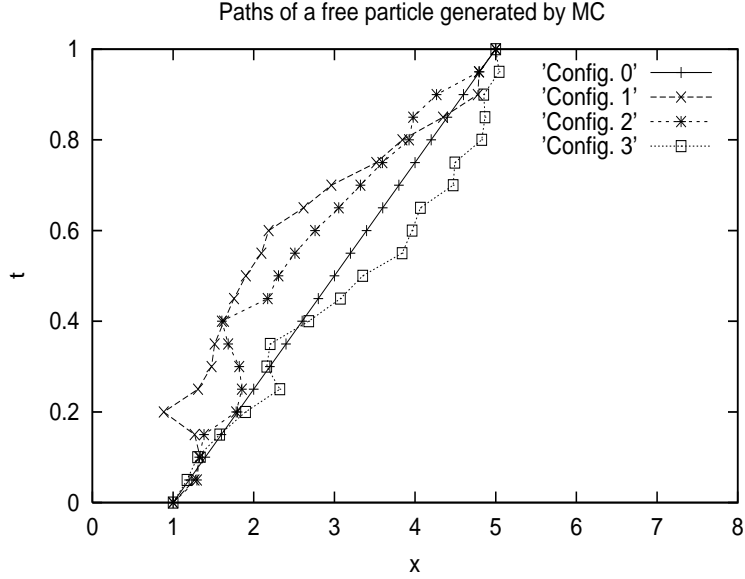, height=10cm, width=8cm,angle=270}}
\caption{Config. 0 stands for the classical path, Config. 1, Config. 2 
and Config. 3 stand the paths generated by MC. 
Here $N_t=20$ (there are 19 time slices between $t_i$ and $t_f$).} 
\label{fig1}
\end{figure}

\subsection{Matrix elements}
\label{Matrix}

As explained above, 
the calculation of the transition matrix elements is an essential ingredient of our method.
The matrix element in the normalized basis is related to 
$\langle x_{n'},t_{f}  \vert  x_{n},t_{i} \rangle$ by 
\begin{eqnarray}
M_{n'n}
&=& 
\langle e_{n'},t_f \vert e_{n},t_i \rangle
= 
\int_{x_n'}^{x_{n'+1}} dx' \int_{x_n}^{x_{n+1}} dx''  ~
{ \langle x',t_{f}  \vert  x'',t_{i} \rangle \over \sqrt{\Delta x_{n'}\Delta x_{n}}}
\nonumber \\
&\approx& 
\sqrt{\Delta x_{n'} \Delta x_{n}} \langle x_{n'},t_{f}  \vert  x_{n},t_{i} \rangle,
\label{ME}
\end{eqnarray}
where for the regular basis, $\Delta x_n = {\rm const.}$,
and for a stochastic basis, $\Delta x_n= 1/(P_{basis}[x_n]N)$. 
$\langle x_{n'},t_{f}  \vert  x_{n},t_{i} \rangle$ 
can be calculated by MC as follows.

\noindent
(a) Discretize the continuous time with time interval $\Delta t=T/N_t$.

\noindent
(b) Generate free configurations $[x]$ 
between $t \in (t_i,t_f)$ obeying the Boltzmann distribution
\begin{eqnarray}
P_0[x] = { \exp(- S_0[x]/ \hbar)  \over
       \int [dx] \exp(- S_0[x]/ \hbar)}\vert_{x_n,t_i}^{x_{n'},t_f}.
\end{eqnarray}
Figure \ref{fig1} shows a sample of free configurations 
generated by MC method with importance sampling.

\noindent
(c) Measure
\begin{eqnarray} 
\langle O_V \rangle=
\int [dx] ~ \exp(- S_V(x)/ \hbar)\vert_{x_n,t_i}^{x_{n'},t_f} 
~~ P_0[x].
\end{eqnarray}
The path integral in Eq. (\ref{ME}) is then
\begin{eqnarray}
\langle x_{n'},t_{f}  \vert  x_{n},t_{i} \rangle &=& \langle O_V \rangle  
\sqrt{m \over 2\pi \hbar T}
 \exp \left[ -{m \over 2 \hbar T} (x_{n'}-x_n)^2 \right].
\end{eqnarray}

\section {Quantum Field Theory}   

\subsection{The Klein-Gordon model}

The main purpose of the algorithm is to study
many body systems and QFT.
As an example, we consider a free scalar field theory:
the Klein-Gordon model in
1+1 dimensions, which has the Hamiltonian
\begin{eqnarray}
H=\int dx {1 \over 2} \left(\pi^2 
+ \nu^2 \left( {\partial \varphi \over \partial x} \right)^2
+\Omega_0^2 \varphi^2 \right).
\end{eqnarray}
Discretizing the theory on a finite string with $N_{\rm{osc}}$ sites
and lattice spacing $a$, 
and making the replacement $x \to x_j= ja$,
$\pi (x) \to a^{-1/2} \Pi(x_j)$,
$\varphi (x) \to a^{-1/2} \phi(x_j)$, 
$\int \to a \sum_{j=1}^{N_{\rm{osc}}}$ and $\nu \to a \Omega$, 
the Hamiltonian becomes 
\begin{eqnarray}
H=\sum_{j=1}^{N_{\rm{osc}}} {1 \over 2} \left( \Pi^2(x_j)
+\Omega^2 (\phi(x_j)-\phi (x_{j+1}))^2+\Omega_0^2 \phi^2 (x_j) \right),
\end{eqnarray}
which describes a chain of $N_{\rm{osc}}$ coupled oscillators\cite{Henley}
with $m=1$ in 1 spatial dimension.  
The field $\phi (x_j)$  
and its conjugate momentum  $\Pi (x_j)={\dot \phi (x_j)}$ 
at the $j$-th site
correspond respectively to the 
displacement  and velocity 
of the $j$-th oscillator. 

In QFT, the initial and final states are defined by
\begin{eqnarray*}
\vert \Phi_i, t_i \rangle=\vert \phi (x_1,t_i), ..., \phi (x_{N_{\rm{osc}}}, t_i) \rangle,
\end{eqnarray*}
\begin{eqnarray}
\vert \Phi_f, t_f \rangle=\vert \phi (x_1,t_f), ..., \phi (x_{N_{\rm{osc}}}, t_f) \rangle.
\end{eqnarray}
The transition amplitude between them is
\begin{eqnarray}
M_{fi} &=& \langle \Phi_{f},t_{f}  \vert  \Phi_{i},t_{i} \rangle =
\langle \Phi_{f}  \vert  e^{-H(t_{f}-t_{i})/\hbar}  \vert  \Phi_{i} \rangle
\nonumber \\
&=& 
\int [d\phi] \exp(- S[\phi]/ \hbar) \vert_{\phi_i,t_i}^{\phi_f,t_f}
\nonumber \\
& = &
\lim_{N_t \to \infty}
\int \prod_{k=1}^{N_t-1} \prod_{j=1}^{N_{\rm{osc}}} 
\left( \sqrt{1 \over 2 \pi \hbar \Delta t} d\phi (x_j,t_k) \right)
\exp(- S[\phi]/ \hbar),
\end{eqnarray}
where the Euclidean action is
\begin{eqnarray}
S 
&=& {\Delta t \over 2} 
\sum_{k=1}^{N_t-1} \sum_{j=1}^{N_{\rm{osc}}} 
\left[
{\left(\phi (x_j,t_{k+1})-\phi (x_j,t_k)\right)^2 \over (\Delta t)^2}
\nonumber \right.\\
&+&
\Omega^2 \left(\phi (x_j,t_k)-\phi (x_{j+1},t_k)\right)^2
+\Omega_0^2 \phi^2(x_j,t_k) 
\bigg].
\end{eqnarray}
For a finite $N_{\rm{osc}}$ and non-zero $\Omega$, 
one has to introduce a boundary condition for $\phi (x_{N_{\rm{osc}}+1})$.
We implement the periodic boundary condition 
$\phi (x_{N_{\rm{osc}}+1})=\phi (x_1)$.
The spectrum is analytically known\cite{Henley}:
\begin{eqnarray} 
E_{n}=\sum_{n_l} ~~ \left( n_l+{1\over 2} \right)  \hbar \omega_l, 
\label{spectrum}
\end{eqnarray}
where
$\omega_l=\sqrt{\Omega_0^2 + 4 \Omega^2 \sin^2 (p_l \Delta x/2)}$, 
and
$n_1,...,n_{N_{\rm{osc}}}= 0,1,\ldots ~$. 
We can also compute the transition amplitude analytically,
using the discrete Fourier transformation
\begin{eqnarray}
\phi (x_j,t) &=& \sum_l {\Delta p \over \sqrt{2\pi}} 
\exp(i p_l x_j) {\tilde \phi} (p_l,t), 
\nonumber \\
{\tilde \phi} (p_l,t)   &=& 
\sum_j {\Delta x \over \sqrt{2\pi}}\exp(-ip_l x_j) \phi (x_j,t),
\end{eqnarray}
where $\Delta x=a$, $\Delta p= 2 \pi / (N_{\rm{osc}} \Delta x)$,
$x_j= [-(N_{\rm{osc}} -1) / 2 + (j-1)]\Delta x$,
$p_l= [-(N_{\rm{osc}} -1) / 2 + (l-1)]\Delta p$,
and $j,l=1,..., N_{\rm{osc}}$.
The result is  
\begin{eqnarray}
M_{fi} &=& 
 \prod_{l=1}^{N_{\rm{osc}}} 
\sqrt{m \omega_l \over 2 \pi \hbar {\rm sinh} (\omega_l T)} 
~
\exp 
\left[
- {\omega_l \Delta p/\Delta x 
\over 2 \hbar {\rm sinh} (\omega_l T)} 
\nonumber \right.\\
& \times &
\left(
  \left( 
\vert \tilde{\phi} (p_l,t_f) \vert^2 
+\vert \tilde{\phi} (p_l,t_i) \vert^2 
  \right) 
{\rm cosh} (\omega_l T)
- 2 {\rm Re} 
  \left( 
\tilde{\phi} (p_l,t_f) \tilde{\phi}^{\star} (p_l,t_i)
  \right)
\right)
\bigg].
\nonumber \\
& &
\label{MKG}
\end{eqnarray}

\subsection{Stochastic basis}
In QFT, a stochastic basis has to be used for the
initial and final states, because in the thermodynamical limit, 
$N_{{\rm osc}} \to \infty$. 
Although the variables $\phi$  are coupled in position space,
we may still use a stochastic basis $[\Phi_n]$, $n=1,...,N$ 
according to the distribution 
with an adjustable parameter $\sigma$
\begin{eqnarray}
P_{basis}[\Phi_n] =
\prod_{j=1}^{N_{\rm{osc}}} {1 \over \sqrt{2\pi} \sigma} 
\exp \left(-{\phi_n^2(x_j) \over 2 \sigma^2} \right).
\end{eqnarray}
The matrix element between the normalized initial
and final states is
\begin{eqnarray}
M_{n'n}
= 
\langle e_{n'},t_f \vert e_{n},t_i \rangle
\approx
\sqrt{\Delta \Phi_{n'} \Delta \Phi_{n}} 
\langle \Phi_{n'},t_{f}  \vert  \Phi_{n},t_{i} \rangle,
\label{MFT}
\end{eqnarray}
where the infinitesimal volumes for the initial and final states
are 
\begin{eqnarray}
\Delta \Phi_{n} &=& d \phi_n (x_1) ... d \phi_n (x_{N_{\rm{osc}}})=
{1 \over P_{basis}[\Phi_n] N},
\nonumber \\
\Delta \Phi_{n'} &=& d \phi_{n'} (x_1) ... d \phi_{n'} (x_{N_{\rm{osc}}})=
{1 \over P_{basis}[\Phi_{n'}] N}.
\label{STO_PHI}
\end{eqnarray}

\section{Results}

We consider $N_{\rm{osc}}=9$, $a=1$, $\Omega=1$, $\Omega_0=2$, 
$m=1$, $\hbar=1$, and $T=2$. 
For the adjustable parameter $\sigma$ in the stochastic basis, 
we choose $\sigma=\sqrt{\hbar {\rm sinh}(\Omega_0 T')/(m \Omega_0)}$ 
with $T'=T$
for simplicity. 
After the stochastic basis with $N=1000$ is generated,  
we obtain the matrix elements $M_{n'n}$ 
using Eqs. (\ref{MKG}), (\ref{MFT}) and (\ref{STO_PHI}). 
(In the future work, we will calculate 
$\langle \Phi_{n'},t_{f}  \vert  \Phi_{n},t_{i} \rangle$
directly by MC).
Then we compute the eigenvalues and eigenvectors 
using the method described in Sect. \ref{EH}.

Table \ref{tab.3} gives a comparison between the spectrum 
from the effective Hamiltonian with the stochastic basis  
and the analytical formular Eq. (\ref{spectrum}) for the first 20 states.
They agree very well. 
This means that an arbitrary choice of Gaussian distribution for
the stochastic basis is good enough. (Of course, one should study 
systematically the dependence of the results on $\sigma$).

\begin{table}[hbt]
\caption{Comparison of the spectrum of the Klein-Gordon model on the lattice,
between the MC Hamiltonian method with a {\it stochastic} basis and the analytic ones.}
\begin{center}
\begin{tabular}{|c|c|c|}
\hline
$n$ & $E_{n}^{\rm{eff}}$  & $E_{n}^{\rm{exact}}$\\
\hline
   1   & 10.904663192168  &  10.944060480668\\
   2   & 12.956830557334  &  12.944060480668\\
   3   & 12.985023578737  &  13.057803869484\\
   4   & 13.044311582647  &  13.057803869484\\
   5   & 13.299967341242  &  13.321601993380\\
   6   & 13.345480638394  &  13.321601993380\\
   7   & 13.552195133687  &  13.589811791733\\
   8   & 13.585794986361  &  13.589811791733\\
   9   & 13.680136748933  &  13.751084748745\\
   10  &  13.744919087477 &   13.751084748745\\
   11  &  14.984737011385 &   14.944060480668\\
   12  &  15.012353803145 &   15.057803869484\\
   13  &  15.057295761044 &   15.057803869484\\
   14  &  15.108904652020 &   15.171547258300\\
   15  &  15.125356713561 &   15.171547258300\\
   16  &  15.187413290039 &   15.171547258300\\
   17  &  15.308536490102 &   15.321601993380\\
   18  &  15.396255686587 &   15.321601993380\\
   19  &  15.420708031412 &   15.435345382196\\
   20  &  15.432823810789 &   15.435345382196\\
\hline
\end{tabular}
\end{center}
\label{tab.3}
\end{table}

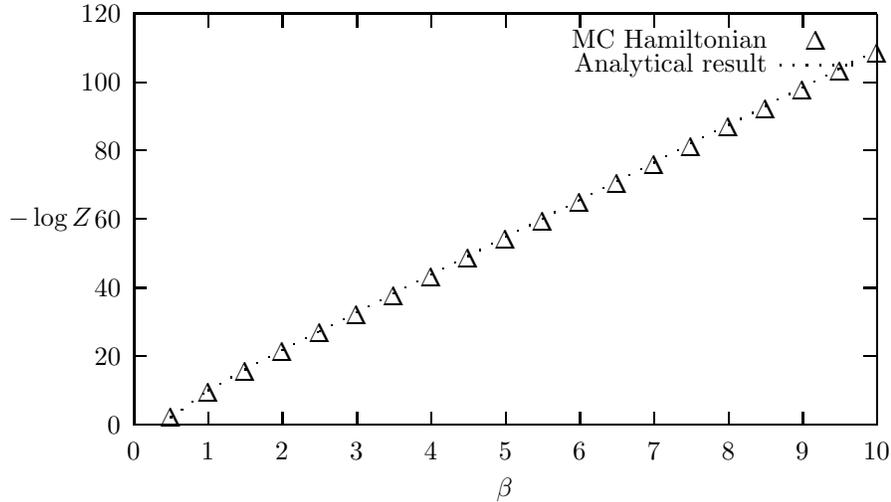
\begin{figure}[htb]
\begin{center}
\setlength{\unitlength}{0.240900pt}
\ifx\plotpoint\undefined\newsavebox{\plotpoint}\fi
\sbox{\plotpoint}{\rule[-0.200pt]{0.400pt}{0.400pt}}%
\begin{picture}(1349,809)(0,0)
\font\gnuplot=cmr10 at 10pt
\gnuplot
\sbox{\plotpoint}{\rule[-0.200pt]{0.400pt}{0.400pt}}%
\put(161.0,123.0){\rule[-0.200pt]{4.818pt}{0.400pt}}
\put(141,123){\makebox(0,0)[r]{0}}
\put(1308.0,123.0){\rule[-0.200pt]{4.818pt}{0.400pt}}
\put(161.0,231.0){\rule[-0.200pt]{4.818pt}{0.400pt}}
\put(141,231){\makebox(0,0)[r]{20}}
\put(1308.0,231.0){\rule[-0.200pt]{4.818pt}{0.400pt}}
\put(161.0,338.0){\rule[-0.200pt]{4.818pt}{0.400pt}}
\put(141,338){\makebox(0,0)[r]{40}}
\put(1308.0,338.0){\rule[-0.200pt]{4.818pt}{0.400pt}}
\put(161.0,446.0){\rule[-0.200pt]{4.818pt}{0.400pt}}
\put(141,446){\makebox(0,0)[r]{60}}
\put(1308.0,446.0){\rule[-0.200pt]{4.818pt}{0.400pt}}
\put(161.0,554.0){\rule[-0.200pt]{4.818pt}{0.400pt}}
\put(141,554){\makebox(0,0)[r]{80}}
\put(1308.0,554.0){\rule[-0.200pt]{4.818pt}{0.400pt}}
\put(161.0,661.0){\rule[-0.200pt]{4.818pt}{0.400pt}}
\put(141,661){\makebox(0,0)[r]{100}}
\put(1308.0,661.0){\rule[-0.200pt]{4.818pt}{0.400pt}}
\put(161.0,769.0){\rule[-0.200pt]{4.818pt}{0.400pt}}
\put(141,769){\makebox(0,0)[r]{120}}
\put(1308.0,769.0){\rule[-0.200pt]{4.818pt}{0.400pt}}
\put(161.0,123.0){\rule[-0.200pt]{0.400pt}{4.818pt}}
\put(161,82){\makebox(0,0){0}}
\put(161.0,749.0){\rule[-0.200pt]{0.400pt}{4.818pt}}
\put(278.0,123.0){\rule[-0.200pt]{0.400pt}{4.818pt}}
\put(278,82){\makebox(0,0){1}}
\put(278.0,749.0){\rule[-0.200pt]{0.400pt}{4.818pt}}
\put(394.0,123.0){\rule[-0.200pt]{0.400pt}{4.818pt}}
\put(394,82){\makebox(0,0){2}}
\put(394.0,749.0){\rule[-0.200pt]{0.400pt}{4.818pt}}
\put(511.0,123.0){\rule[-0.200pt]{0.400pt}{4.818pt}}
\put(511,82){\makebox(0,0){3}}
\put(511.0,749.0){\rule[-0.200pt]{0.400pt}{4.818pt}}
\put(628.0,123.0){\rule[-0.200pt]{0.400pt}{4.818pt}}
\put(628,82){\makebox(0,0){4}}
\put(628.0,749.0){\rule[-0.200pt]{0.400pt}{4.818pt}}
\put(745.0,123.0){\rule[-0.200pt]{0.400pt}{4.818pt}}
\put(745,82){\makebox(0,0){5}}
\put(745.0,749.0){\rule[-0.200pt]{0.400pt}{4.818pt}}
\put(861.0,123.0){\rule[-0.200pt]{0.400pt}{4.818pt}}
\put(861,82){\makebox(0,0){6}}
\put(861.0,749.0){\rule[-0.200pt]{0.400pt}{4.818pt}}
\put(978.0,123.0){\rule[-0.200pt]{0.400pt}{4.818pt}}
\put(978,82){\makebox(0,0){7}}
\put(978.0,749.0){\rule[-0.200pt]{0.400pt}{4.818pt}}
\put(1095.0,123.0){\rule[-0.200pt]{0.400pt}{4.818pt}}
\put(1095,82){\makebox(0,0){8}}
\put(1095.0,749.0){\rule[-0.200pt]{0.400pt}{4.818pt}}
\put(1211.0,123.0){\rule[-0.200pt]{0.400pt}{4.818pt}}
\put(1211,82){\makebox(0,0){9}}
\put(1211.0,749.0){\rule[-0.200pt]{0.400pt}{4.818pt}}
\put(1328.0,123.0){\rule[-0.200pt]{0.400pt}{4.818pt}}
\put(1328,82){\makebox(0,0){10}}
\put(1328.0,749.0){\rule[-0.200pt]{0.400pt}{4.818pt}}
\put(161.0,123.0){\rule[-0.200pt]{281.130pt}{0.400pt}}
\put(1328.0,123.0){\rule[-0.200pt]{0.400pt}{155.621pt}}
\put(161.0,769.0){\rule[-0.200pt]{281.130pt}{0.400pt}}
\put(30,446){\makebox(0,0){$-\log Z$}}
\put(744,21){\makebox(0,0){$\beta$}}
\put(161.0,123.0){\rule[-0.200pt]{0.400pt}{155.621pt}}
\put(1156,729){\makebox(0,0)[r]{MC Hamiltonian}}
\put(219,138){\raisebox{-.8pt}{\makebox(0,0){$\Delta$}}}
\put(278,177){\raisebox{-.8pt}{\makebox(0,0){$\Delta$}}}
\put(336,210){\raisebox{-.8pt}{\makebox(0,0){$\Delta$}}}
\put(394,240){\raisebox{-.8pt}{\makebox(0,0){$\Delta$}}}
\put(453,270){\raisebox{-.8pt}{\makebox(0,0){$\Delta$}}}
\put(511,299){\raisebox{-.8pt}{\makebox(0,0){$\Delta$}}}
\put(569,328){\raisebox{-.8pt}{\makebox(0,0){$\Delta$}}}
\put(628,358){\raisebox{-.8pt}{\makebox(0,0){$\Delta$}}}
\put(686,387){\raisebox{-.8pt}{\makebox(0,0){$\Delta$}}}
\put(745,417){\raisebox{-.8pt}{\makebox(0,0){$\Delta$}}}
\put(803,446){\raisebox{-.8pt}{\makebox(0,0){$\Delta$}}}
\put(861,475){\raisebox{-.8pt}{\makebox(0,0){$\Delta$}}}
\put(920,505){\raisebox{-.8pt}{\makebox(0,0){$\Delta$}}}
\put(978,534){\raisebox{-.8pt}{\makebox(0,0){$\Delta$}}}
\put(1036,563){\raisebox{-.8pt}{\makebox(0,0){$\Delta$}}}
\put(1095,593){\raisebox{-.8pt}{\makebox(0,0){$\Delta$}}}
\put(1153,622){\raisebox{-.8pt}{\makebox(0,0){$\Delta$}}}
\put(1211,651){\raisebox{-.8pt}{\makebox(0,0){$\Delta$}}}
\put(1270,681){\raisebox{-.8pt}{\makebox(0,0){$\Delta$}}}
\put(1328,710){\raisebox{-.8pt}{\makebox(0,0){$\Delta$}}}
\put(1232,729){\raisebox{-.8pt}{\makebox(0,0){$\Delta$}}}
\put(1156,688){\makebox(0,0)[r]{Analytical result}}
\multiput(1176,688)(20.756,0.000){6}{\usebox{\plotpoint}}
\put(1288,688){\usebox{\plotpoint}}
\put(219,135){\usebox{\plotpoint}}
\multiput(219,135)(16.909,12.037){4}{\usebox{\plotpoint}}
\multiput(278,177)(18.040,10.264){3}{\usebox{\plotpoint}}
\multiput(336,210)(18.435,9.536){3}{\usebox{\plotpoint}}
\multiput(394,240)(18.501,9.407){4}{\usebox{\plotpoint}}
\multiput(453,270)(18.435,9.536){3}{\usebox{\plotpoint}}
\multiput(511,300)(18.564,9.282){3}{\usebox{\plotpoint}}
\multiput(569,329)(18.501,9.407){3}{\usebox{\plotpoint}}
\multiput(628,359)(18.564,9.282){3}{\usebox{\plotpoint}}
\multiput(686,388)(18.501,9.407){3}{\usebox{\plotpoint}}
\multiput(745,418)(18.564,9.282){3}{\usebox{\plotpoint}}
\multiput(803,447)(18.564,9.282){4}{\usebox{\plotpoint}}
\multiput(861,476)(18.501,9.407){3}{\usebox{\plotpoint}}
\multiput(920,506)(18.564,9.282){3}{\usebox{\plotpoint}}
\multiput(978,535)(18.435,9.536){3}{\usebox{\plotpoint}}
\multiput(1036,565)(18.627,9.156){3}{\usebox{\plotpoint}}
\multiput(1095,594)(18.435,9.536){3}{\usebox{\plotpoint}}
\multiput(1153,624)(18.564,9.282){3}{\usebox{\plotpoint}}
\multiput(1211,653)(18.501,9.407){4}{\usebox{\plotpoint}}
\multiput(1270,683)(18.564,9.282){3}{\usebox{\plotpoint}}
\put(1328,712){\usebox{\plotpoint}}
\end{picture}
\end{center}
\caption{Free energy of the Klein-Gordon model on a 1+1 dimensional lattice.} 
\label{fig.2}
\end{figure}

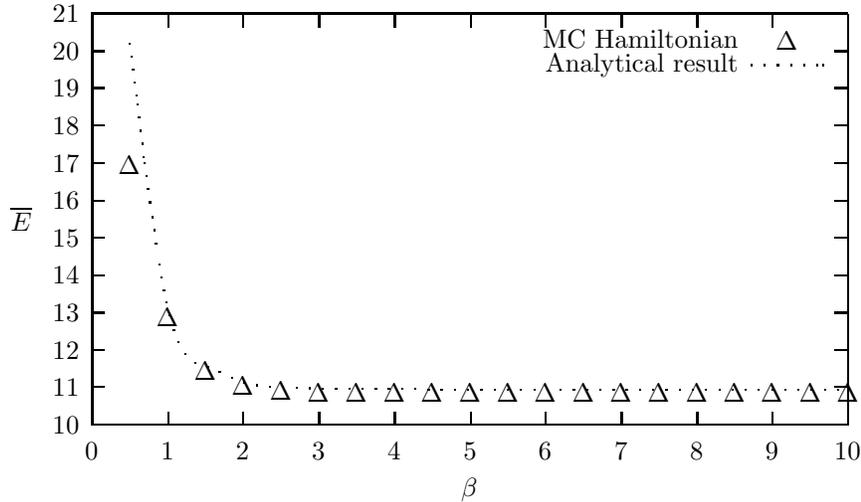
\begin{figure}[htb]
\begin{center}
\setlength{\unitlength}{0.240900pt}
\ifx\plotpoint\undefined\newsavebox{\plotpoint}\fi
\sbox{\plotpoint}{\rule[-0.200pt]{0.400pt}{0.400pt}}%
\begin{picture}(1349,809)(0,0)
\font\gnuplot=cmr10 at 10pt
\gnuplot
\sbox{\plotpoint}{\rule[-0.200pt]{0.400pt}{0.400pt}}%
\put(141.0,123.0){\rule[-0.200pt]{4.818pt}{0.400pt}}
\put(121,123){\makebox(0,0)[r]{10}}
\put(1308.0,123.0){\rule[-0.200pt]{4.818pt}{0.400pt}}
\put(141.0,182.0){\rule[-0.200pt]{4.818pt}{0.400pt}}
\put(121,182){\makebox(0,0)[r]{11}}
\put(1308.0,182.0){\rule[-0.200pt]{4.818pt}{0.400pt}}
\put(141.0,240.0){\rule[-0.200pt]{4.818pt}{0.400pt}}
\put(121,240){\makebox(0,0)[r]{12}}
\put(1308.0,240.0){\rule[-0.200pt]{4.818pt}{0.400pt}}
\put(141.0,299.0){\rule[-0.200pt]{4.818pt}{0.400pt}}
\put(121,299){\makebox(0,0)[r]{13}}
\put(1308.0,299.0){\rule[-0.200pt]{4.818pt}{0.400pt}}
\put(141.0,358.0){\rule[-0.200pt]{4.818pt}{0.400pt}}
\put(121,358){\makebox(0,0)[r]{14}}
\put(1308.0,358.0){\rule[-0.200pt]{4.818pt}{0.400pt}}
\put(141.0,417.0){\rule[-0.200pt]{4.818pt}{0.400pt}}
\put(121,417){\makebox(0,0)[r]{15}}
\put(1308.0,417.0){\rule[-0.200pt]{4.818pt}{0.400pt}}
\put(141.0,475.0){\rule[-0.200pt]{4.818pt}{0.400pt}}
\put(121,475){\makebox(0,0)[r]{16}}
\put(1308.0,475.0){\rule[-0.200pt]{4.818pt}{0.400pt}}
\put(141.0,534.0){\rule[-0.200pt]{4.818pt}{0.400pt}}
\put(121,534){\makebox(0,0)[r]{17}}
\put(1308.0,534.0){\rule[-0.200pt]{4.818pt}{0.400pt}}
\put(141.0,593.0){\rule[-0.200pt]{4.818pt}{0.400pt}}
\put(121,593){\makebox(0,0)[r]{18}}
\put(1308.0,593.0){\rule[-0.200pt]{4.818pt}{0.400pt}}
\put(141.0,652.0){\rule[-0.200pt]{4.818pt}{0.400pt}}
\put(121,652){\makebox(0,0)[r]{19}}
\put(1308.0,652.0){\rule[-0.200pt]{4.818pt}{0.400pt}}
\put(141.0,710.0){\rule[-0.200pt]{4.818pt}{0.400pt}}
\put(121,710){\makebox(0,0)[r]{20}}
\put(1308.0,710.0){\rule[-0.200pt]{4.818pt}{0.400pt}}
\put(141.0,769.0){\rule[-0.200pt]{4.818pt}{0.400pt}}
\put(121,769){\makebox(0,0)[r]{21}}
\put(1308.0,769.0){\rule[-0.200pt]{4.818pt}{0.400pt}}
\put(141.0,123.0){\rule[-0.200pt]{0.400pt}{4.818pt}}
\put(141,82){\makebox(0,0){0}}
\put(141.0,749.0){\rule[-0.200pt]{0.400pt}{4.818pt}}
\put(260.0,123.0){\rule[-0.200pt]{0.400pt}{4.818pt}}
\put(260,82){\makebox(0,0){1}}
\put(260.0,749.0){\rule[-0.200pt]{0.400pt}{4.818pt}}
\put(378.0,123.0){\rule[-0.200pt]{0.400pt}{4.818pt}}
\put(378,82){\makebox(0,0){2}}
\put(378.0,749.0){\rule[-0.200pt]{0.400pt}{4.818pt}}
\put(497.0,123.0){\rule[-0.200pt]{0.400pt}{4.818pt}}
\put(497,82){\makebox(0,0){3}}
\put(497.0,749.0){\rule[-0.200pt]{0.400pt}{4.818pt}}
\put(616.0,123.0){\rule[-0.200pt]{0.400pt}{4.818pt}}
\put(616,82){\makebox(0,0){4}}
\put(616.0,749.0){\rule[-0.200pt]{0.400pt}{4.818pt}}
\put(735.0,123.0){\rule[-0.200pt]{0.400pt}{4.818pt}}
\put(735,82){\makebox(0,0){5}}
\put(735.0,749.0){\rule[-0.200pt]{0.400pt}{4.818pt}}
\put(853.0,123.0){\rule[-0.200pt]{0.400pt}{4.818pt}}
\put(853,82){\makebox(0,0){6}}
\put(853.0,749.0){\rule[-0.200pt]{0.400pt}{4.818pt}}
\put(972.0,123.0){\rule[-0.200pt]{0.400pt}{4.818pt}}
\put(972,82){\makebox(0,0){7}}
\put(972.0,749.0){\rule[-0.200pt]{0.400pt}{4.818pt}}
\put(1091.0,123.0){\rule[-0.200pt]{0.400pt}{4.818pt}}
\put(1091,82){\makebox(0,0){8}}
\put(1091.0,749.0){\rule[-0.200pt]{0.400pt}{4.818pt}}
\put(1209.0,123.0){\rule[-0.200pt]{0.400pt}{4.818pt}}
\put(1209,82){\makebox(0,0){9}}
\put(1209.0,749.0){\rule[-0.200pt]{0.400pt}{4.818pt}}
\put(1328.0,123.0){\rule[-0.200pt]{0.400pt}{4.818pt}}
\put(1328,82){\makebox(0,0){10}}
\put(1328.0,749.0){\rule[-0.200pt]{0.400pt}{4.818pt}}
\put(141.0,123.0){\rule[-0.200pt]{285.948pt}{0.400pt}}
\put(1328.0,123.0){\rule[-0.200pt]{0.400pt}{155.621pt}}
\put(141.0,769.0){\rule[-0.200pt]{285.948pt}{0.400pt}}
\put(30,446){\makebox(0,0){$\overline{E}$}}
\put(734,21){\makebox(0,0){$\beta$}}
\put(141.0,123.0){\rule[-0.200pt]{0.400pt}{155.621pt}}
\put(1156,729){\makebox(0,0)[r]{MC Hamiltonian}}
\put(200,534){\raisebox{-.8pt}{\makebox(0,0){$\Delta$}}}
\put(260,296){\raisebox{-.8pt}{\makebox(0,0){$\Delta$}}}
\put(319,211){\raisebox{-.8pt}{\makebox(0,0){$\Delta$}}}
\put(378,187){\raisebox{-.8pt}{\makebox(0,0){$\Delta$}}}
\put(438,179){\raisebox{-.8pt}{\makebox(0,0){$\Delta$}}}
\put(497,177){\raisebox{-.8pt}{\makebox(0,0){$\Delta$}}}
\put(556,176){\raisebox{-.8pt}{\makebox(0,0){$\Delta$}}}
\put(616,176){\raisebox{-.8pt}{\makebox(0,0){$\Delta$}}}
\put(675,176){\raisebox{-.8pt}{\makebox(0,0){$\Delta$}}}
\put(735,176){\raisebox{-.8pt}{\makebox(0,0){$\Delta$}}}
\put(794,176){\raisebox{-.8pt}{\makebox(0,0){$\Delta$}}}
\put(853,176){\raisebox{-.8pt}{\makebox(0,0){$\Delta$}}}
\put(913,176){\raisebox{-.8pt}{\makebox(0,0){$\Delta$}}}
\put(972,176){\raisebox{-.8pt}{\makebox(0,0){$\Delta$}}}
\put(1031,176){\raisebox{-.8pt}{\makebox(0,0){$\Delta$}}}
\put(1091,176){\raisebox{-.8pt}{\makebox(0,0){$\Delta$}}}
\put(1150,176){\raisebox{-.8pt}{\makebox(0,0){$\Delta$}}}
\put(1209,176){\raisebox{-.8pt}{\makebox(0,0){$\Delta$}}}
\put(1269,176){\raisebox{-.8pt}{\makebox(0,0){$\Delta$}}}
\put(1328,176){\raisebox{-.8pt}{\makebox(0,0){$\Delta$}}}
\put(1232,729){\raisebox{-.8pt}{\makebox(0,0){$\Delta$}}}
\put(1156,688){\makebox(0,0)[r]{Analytical result}}
\multiput(1176,688)(20.756,0.000){6}{\usebox{\plotpoint}}
\put(1288,688){\usebox{\plotpoint}}
\put(200,722){\usebox{\plotpoint}}
\multiput(200,722)(2.574,-20.595){5}{\usebox{\plotpoint}}
\multiput(212,626)(2.464,-20.609){5}{\usebox{\plotpoint}}
\multiput(223,534)(2.901,-20.552){4}{\usebox{\plotpoint}}
\multiput(235,449)(3.052,-20.530){3}{\usebox{\plotpoint}}
\multiput(246,375)(3.743,-20.415){3}{\usebox{\plotpoint}}
\multiput(257,315)(5.579,-19.992){2}{\usebox{\plotpoint}}
\multiput(269,272)(7.589,-19.318){2}{\usebox{\plotpoint}}
\put(287.00,233.83){\usebox{\plotpoint}}
\put(302.39,220.41){\usebox{\plotpoint}}
\put(321.89,213.37){\usebox{\plotpoint}}
\put(341.14,205.74){\usebox{\plotpoint}}
\put(359.95,197.02){\usebox{\plotpoint}}
\put(379.17,189.28){\usebox{\plotpoint}}
\put(399.45,185.01){\usebox{\plotpoint}}
\put(420.06,182.72){\usebox{\plotpoint}}
\put(440.78,181.93){\usebox{\plotpoint}}
\put(461.49,181.00){\usebox{\plotpoint}}
\put(482.21,180.00){\usebox{\plotpoint}}
\put(502.94,179.46){\usebox{\plotpoint}}
\put(523.67,179.00){\usebox{\plotpoint}}
\put(544.43,179.00){\usebox{\plotpoint}}
\put(565.19,179.00){\usebox{\plotpoint}}
\put(585.94,179.00){\usebox{\plotpoint}}
\put(606.70,179.00){\usebox{\plotpoint}}
\put(627.45,179.00){\usebox{\plotpoint}}
\put(648.19,178.71){\usebox{\plotpoint}}
\put(668.92,178.00){\usebox{\plotpoint}}
\put(689.67,178.00){\usebox{\plotpoint}}
\put(710.43,178.00){\usebox{\plotpoint}}
\put(731.18,178.00){\usebox{\plotpoint}}
\put(751.94,178.00){\usebox{\plotpoint}}
\put(772.69,178.00){\usebox{\plotpoint}}
\put(793.45,178.00){\usebox{\plotpoint}}
\put(814.21,178.00){\usebox{\plotpoint}}
\put(834.96,178.00){\usebox{\plotpoint}}
\put(855.72,178.00){\usebox{\plotpoint}}
\put(876.47,178.00){\usebox{\plotpoint}}
\put(897.23,178.00){\usebox{\plotpoint}}
\put(917.98,178.00){\usebox{\plotpoint}}
\put(938.74,178.00){\usebox{\plotpoint}}
\put(959.49,178.00){\usebox{\plotpoint}}
\put(980.25,178.00){\usebox{\plotpoint}}
\put(1001.01,178.00){\usebox{\plotpoint}}
\put(1021.76,178.00){\usebox{\plotpoint}}
\put(1042.52,178.00){\usebox{\plotpoint}}
\put(1063.27,178.00){\usebox{\plotpoint}}
\put(1084.03,178.00){\usebox{\plotpoint}}
\put(1104.78,178.00){\usebox{\plotpoint}}
\put(1125.54,178.00){\usebox{\plotpoint}}
\put(1146.29,178.00){\usebox{\plotpoint}}
\put(1167.05,178.00){\usebox{\plotpoint}}
\put(1187.80,178.00){\usebox{\plotpoint}}
\put(1208.56,178.00){\usebox{\plotpoint}}
\put(1229.32,178.00){\usebox{\plotpoint}}
\put(1250.07,178.00){\usebox{\plotpoint}}
\put(1270.83,178.00){\usebox{\plotpoint}}
\put(1291.58,178.00){\usebox{\plotpoint}}
\put(1312.34,178.00){\usebox{\plotpoint}}
\put(1328,178){\usebox{\plotpoint}}
\end{picture}
\end{center}
\caption{Average energy of the Klein-Gordon model on a 1+1 dimensional lattice.} 
\label{fig.3}
\end{figure}

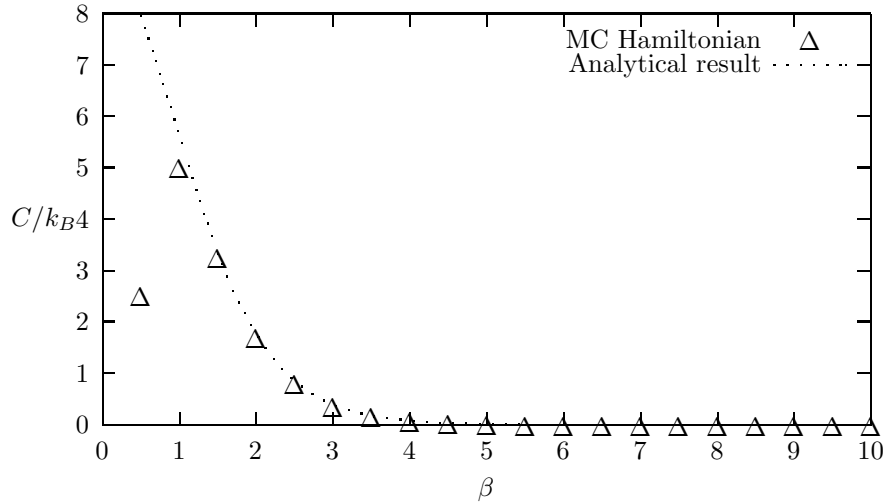
\begin{figure}[htb]
\begin{center}
\setlength{\unitlength}{0.240900pt}
\ifx\plotpoint\undefined\newsavebox{\plotpoint}\fi
\sbox{\plotpoint}{\rule[-0.200pt]{0.400pt}{0.400pt}}%
\begin{picture}(1349,809)(0,0)
\font\gnuplot=cmr10 at 10pt
\gnuplot
\sbox{\plotpoint}{\rule[-0.200pt]{0.400pt}{0.400pt}}%
\put(121.0,123.0){\rule[-0.200pt]{4.818pt}{0.400pt}}
\put(101,123){\makebox(0,0)[r]{0}}
\put(1308.0,123.0){\rule[-0.200pt]{4.818pt}{0.400pt}}
\put(121.0,204.0){\rule[-0.200pt]{4.818pt}{0.400pt}}
\put(101,204){\makebox(0,0)[r]{1}}
\put(1308.0,204.0){\rule[-0.200pt]{4.818pt}{0.400pt}}
\put(121.0,285.0){\rule[-0.200pt]{4.818pt}{0.400pt}}
\put(101,285){\makebox(0,0)[r]{2}}
\put(1308.0,285.0){\rule[-0.200pt]{4.818pt}{0.400pt}}
\put(121.0,365.0){\rule[-0.200pt]{4.818pt}{0.400pt}}
\put(101,365){\makebox(0,0)[r]{3}}
\put(1308.0,365.0){\rule[-0.200pt]{4.818pt}{0.400pt}}
\put(121.0,446.0){\rule[-0.200pt]{4.818pt}{0.400pt}}
\put(101,446){\makebox(0,0)[r]{4}}
\put(1308.0,446.0){\rule[-0.200pt]{4.818pt}{0.400pt}}
\put(121.0,527.0){\rule[-0.200pt]{4.818pt}{0.400pt}}
\put(101,527){\makebox(0,0)[r]{5}}
\put(1308.0,527.0){\rule[-0.200pt]{4.818pt}{0.400pt}}
\put(121.0,608.0){\rule[-0.200pt]{4.818pt}{0.400pt}}
\put(101,608){\makebox(0,0)[r]{6}}
\put(1308.0,608.0){\rule[-0.200pt]{4.818pt}{0.400pt}}
\put(121.0,688.0){\rule[-0.200pt]{4.818pt}{0.400pt}}
\put(101,688){\makebox(0,0)[r]{7}}
\put(1308.0,688.0){\rule[-0.200pt]{4.818pt}{0.400pt}}
\put(121.0,769.0){\rule[-0.200pt]{4.818pt}{0.400pt}}
\put(101,769){\makebox(0,0)[r]{8}}
\put(1308.0,769.0){\rule[-0.200pt]{4.818pt}{0.400pt}}
\put(121.0,123.0){\rule[-0.200pt]{0.400pt}{4.818pt}}
\put(121,82){\makebox(0,0){0}}
\put(121.0,749.0){\rule[-0.200pt]{0.400pt}{4.818pt}}
\put(242.0,123.0){\rule[-0.200pt]{0.400pt}{4.818pt}}
\put(242,82){\makebox(0,0){1}}
\put(242.0,749.0){\rule[-0.200pt]{0.400pt}{4.818pt}}
\put(362.0,123.0){\rule[-0.200pt]{0.400pt}{4.818pt}}
\put(362,82){\makebox(0,0){2}}
\put(362.0,749.0){\rule[-0.200pt]{0.400pt}{4.818pt}}
\put(483.0,123.0){\rule[-0.200pt]{0.400pt}{4.818pt}}
\put(483,82){\makebox(0,0){3}}
\put(483.0,749.0){\rule[-0.200pt]{0.400pt}{4.818pt}}
\put(604.0,123.0){\rule[-0.200pt]{0.400pt}{4.818pt}}
\put(604,82){\makebox(0,0){4}}
\put(604.0,749.0){\rule[-0.200pt]{0.400pt}{4.818pt}}
\put(725.0,123.0){\rule[-0.200pt]{0.400pt}{4.818pt}}
\put(725,82){\makebox(0,0){5}}
\put(725.0,749.0){\rule[-0.200pt]{0.400pt}{4.818pt}}
\put(845.0,123.0){\rule[-0.200pt]{0.400pt}{4.818pt}}
\put(845,82){\makebox(0,0){6}}
\put(845.0,749.0){\rule[-0.200pt]{0.400pt}{4.818pt}}
\put(966.0,123.0){\rule[-0.200pt]{0.400pt}{4.818pt}}
\put(966,82){\makebox(0,0){7}}
\put(966.0,749.0){\rule[-0.200pt]{0.400pt}{4.818pt}}
\put(1087.0,123.0){\rule[-0.200pt]{0.400pt}{4.818pt}}
\put(1087,82){\makebox(0,0){8}}
\put(1087.0,749.0){\rule[-0.200pt]{0.400pt}{4.818pt}}
\put(1207.0,123.0){\rule[-0.200pt]{0.400pt}{4.818pt}}
\put(1207,82){\makebox(0,0){9}}
\put(1207.0,749.0){\rule[-0.200pt]{0.400pt}{4.818pt}}
\put(1328.0,123.0){\rule[-0.200pt]{0.400pt}{4.818pt}}
\put(1328,82){\makebox(0,0){10}}
\put(1328.0,749.0){\rule[-0.200pt]{0.400pt}{4.818pt}}
\put(121.0,123.0){\rule[-0.200pt]{290.766pt}{0.400pt}}
\put(1328.0,123.0){\rule[-0.200pt]{0.400pt}{155.621pt}}
\put(121.0,769.0){\rule[-0.200pt]{290.766pt}{0.400pt}}
\put(30,446){\makebox(0,0){$C/k_B$}}
\put(724,21){\makebox(0,0){$\beta$}}
\put(121.0,123.0){\rule[-0.200pt]{0.400pt}{155.621pt}}
\put(1156,729){\makebox(0,0)[r]{MC Hamiltonian}}
\put(181,327){\raisebox{-.8pt}{\makebox(0,0){$\Delta$}}}
\put(242,528){\raisebox{-.8pt}{\makebox(0,0){$\Delta$}}}
\put(302,386){\raisebox{-.8pt}{\makebox(0,0){$\Delta$}}}
\put(362,261){\raisebox{-.8pt}{\makebox(0,0){$\Delta$}}}
\put(423,189){\raisebox{-.8pt}{\makebox(0,0){$\Delta$}}}
\put(483,153){\raisebox{-.8pt}{\makebox(0,0){$\Delta$}}}
\put(543,136){\raisebox{-.8pt}{\makebox(0,0){$\Delta$}}}
\put(604,129){\raisebox{-.8pt}{\makebox(0,0){$\Delta$}}}
\put(664,125){\raisebox{-.8pt}{\makebox(0,0){$\Delta$}}}
\put(725,124){\raisebox{-.8pt}{\makebox(0,0){$\Delta$}}}
\put(785,123){\raisebox{-.8pt}{\makebox(0,0){$\Delta$}}}
\put(845,123){\raisebox{-.8pt}{\makebox(0,0){$\Delta$}}}
\put(906,123){\raisebox{-.8pt}{\makebox(0,0){$\Delta$}}}
\put(966,123){\raisebox{-.8pt}{\makebox(0,0){$\Delta$}}}
\put(1026,123){\raisebox{-.8pt}{\makebox(0,0){$\Delta$}}}
\put(1087,123){\raisebox{-.8pt}{\makebox(0,0){$\Delta$}}}
\put(1147,123){\raisebox{-.8pt}{\makebox(0,0){$\Delta$}}}
\put(1207,123){\raisebox{-.8pt}{\makebox(0,0){$\Delta$}}}
\put(1268,123){\raisebox{-.8pt}{\makebox(0,0){$\Delta$}}}
\put(1328,123){\raisebox{-.8pt}{\makebox(0,0){$\Delta$}}}
\put(1232,729){\raisebox{-.8pt}{\makebox(0,0){$\Delta$}}}
\put(1156,688){\makebox(0,0)[r]{Analytical result}}
\multiput(1176,688)(20.756,0.000){6}{\usebox{\plotpoint}}
\put(1288,688){\usebox{\plotpoint}}
\put(181,766){\usebox{\plotpoint}}
\multiput(181,766)(6.563,-19.690){2}{\usebox{\plotpoint}}
\multiput(193,730)(6.563,-19.690){2}{\usebox{\plotpoint}}
\multiput(205,694)(6.065,-19.850){2}{\usebox{\plotpoint}}
\multiput(216,658)(6.563,-19.690){2}{\usebox{\plotpoint}}
\multiput(228,622)(5.915,-19.895){2}{\usebox{\plotpoint}}
\multiput(239,585)(6.403,-19.743){2}{\usebox{\plotpoint}}
\put(256.72,528.75){\usebox{\plotpoint}}
\multiput(262,511)(6.403,-19.743){2}{\usebox{\plotpoint}}
\multiput(274,474)(6.732,-19.634){2}{\usebox{\plotpoint}}
\multiput(286,439)(6.563,-19.690){2}{\usebox{\plotpoint}}
\put(302.81,391.00){\usebox{\plotpoint}}
\multiput(309,375)(7.589,-19.318){2}{\usebox{\plotpoint}}
\put(326.30,333.35){\usebox{\plotpoint}}
\put(335.31,314.66){\usebox{\plotpoint}}
\multiput(344,298)(9.631,-18.386){2}{\usebox{\plotpoint}}
\put(365.56,260.29){\usebox{\plotpoint}}
\put(376.81,242.84){\usebox{\plotpoint}}
\put(389.59,226.51){\usebox{\plotpoint}}
\put(402.63,210.37){\usebox{\plotpoint}}
\put(417.67,196.10){\usebox{\plotpoint}}
\put(433.68,182.90){\usebox{\plotpoint}}
\put(450.82,171.21){\usebox{\plotpoint}}
\put(468.89,161.05){\usebox{\plotpoint}}
\put(487.83,152.57){\usebox{\plotpoint}}
\put(507.37,145.63){\usebox{\plotpoint}}
\put(527.45,140.39){\usebox{\plotpoint}}
\put(547.74,136.06){\usebox{\plotpoint}}
\put(568.25,133.23){\usebox{\plotpoint}}
\put(588.83,130.67){\usebox{\plotpoint}}
\put(609.43,129.00){\usebox{\plotpoint}}
\put(630.11,127.24){\usebox{\plotpoint}}
\put(650.82,126.00){\usebox{\plotpoint}}
\put(671.53,125.00){\usebox{\plotpoint}}
\put(692.28,124.89){\usebox{\plotpoint}}
\put(713.00,124.00){\usebox{\plotpoint}}
\put(733.75,124.00){\usebox{\plotpoint}}
\put(754.51,124.00){\usebox{\plotpoint}}
\put(775.25,123.73){\usebox{\plotpoint}}
\put(795.98,123.00){\usebox{\plotpoint}}
\put(816.73,123.00){\usebox{\plotpoint}}
\put(837.49,123.00){\usebox{\plotpoint}}
\put(858.25,123.00){\usebox{\plotpoint}}
\put(879.00,123.00){\usebox{\plotpoint}}
\put(899.76,123.00){\usebox{\plotpoint}}
\put(920.51,123.00){\usebox{\plotpoint}}
\put(941.27,123.00){\usebox{\plotpoint}}
\put(962.02,123.00){\usebox{\plotpoint}}
\put(982.78,123.00){\usebox{\plotpoint}}
\put(1003.53,123.00){\usebox{\plotpoint}}
\put(1024.29,123.00){\usebox{\plotpoint}}
\put(1045.05,123.00){\usebox{\plotpoint}}
\put(1065.80,123.00){\usebox{\plotpoint}}
\put(1086.56,123.00){\usebox{\plotpoint}}
\put(1107.31,123.00){\usebox{\plotpoint}}
\put(1128.07,123.00){\usebox{\plotpoint}}
\put(1148.82,123.00){\usebox{\plotpoint}}
\put(1169.58,123.00){\usebox{\plotpoint}}
\put(1190.33,123.00){\usebox{\plotpoint}}
\put(1211.09,123.00){\usebox{\plotpoint}}
\put(1231.84,123.00){\usebox{\plotpoint}}
\put(1252.60,123.00){\usebox{\plotpoint}}
\put(1273.36,123.00){\usebox{\plotpoint}}
\put(1294.11,123.00){\usebox{\plotpoint}}
\put(1314.87,123.00){\usebox{\plotpoint}}
\put(1328,123){\usebox{\plotpoint}}
\end{picture}
\end{center}
\caption{Specific heat over $k_B$ of the Klein-Gordon model on a 1+1 dimensional lattice.} 
\label{fig.4}
\end{figure}

We have also computed thermodynamical quantities 
such as the partition function $Z$,
average energy $\overline{E}$ and specific heat $C$. 
The analytical results are
\begin{eqnarray}
Z(\beta) &=& {\rm Tr} \left( \exp \left( -\beta H \right) \right)
=\prod_{l=1}^{N_{\rm{osc}}} {1 \over 2 {\rm sinh} \left( \beta \hbar \omega_l/2 \right)},
\nonumber \\
\overline{E}(\beta) &=& {1 \over Z} {\rm Tr} \left(H \exp \left( -\beta H \right) \right)
= - {\partial \log Z \over \partial \beta}
= \sum_{l=1}^{N_{\rm{osc}}} {\hbar \omega_l \over 2} \coth \left( \beta \hbar \omega_l/2 \right),
\nonumber \\
C(\beta) &=& k_B {\partial \overline{E} \over \partial {\cal T}}
= -k_B \beta^2 {\partial \overline{E} \over \partial \beta} 
= k_B \sum_{l=1}^{N_{\rm{osc}}} 
\left( {\beta \hbar \omega_l/2 \over 2 {\rm sinh} 
\left( \beta \hbar \omega_l/2 \right)} \right)^2 ,
\end{eqnarray}
where $\beta=T/\hbar$, the temperature ${\cal T} = 1/(\beta k_B)$, and $k_B$ is the Boltzmann constant.  
Since we have approximated $H$ by $H_{\rm{eff}}$, 
we can express those thermodynamical observables  
via the eigenvalues of the effective Hamiltonian
\begin{eqnarray}
Z^{\rm{eff}}(\beta) &=& \sum_{n=1}^{N}e^{-\beta E_{n}^{\rm{eff}}},
\nonumber \\
\overline{E}^{\rm{eff}}(\beta) &=& \sum_{n=1}^{N}
{E_{n}^{\rm{eff}} {\rm e}^{-\beta E_{n}^{\rm{eff}}} \over Z^{\rm{eff}}(\beta)},
\nonumber \\
C^{\rm{eff}}(\beta) &=& k_B{\beta}^2\left(\sum_{n=1}^{N}
{(E_{n}^{\rm{eff}})^2e^{-\beta E_{n}^{\rm{eff}}} \over
Z^{\rm{eff}}(\beta)}-\left(\overline{E}^{\rm{eff}}(\beta)\right)^2 \right).
\end{eqnarray}
Since this is a static system, the eigenvalues should in principle not vary with $\beta$, which is also
confirmed numerically within statistical errors when $\beta$ and $N$ are not too small. 
Using this assumption and
the spectrum at $\beta=2$, we  obtain the thermodynamical quantities for other $\beta$.
The results for the free energy, average energy and specific heat 
as a function of $\beta$
are shown in Figs. \ref{fig.2},  \ref{fig.3} and  \ref{fig.4}.
Again, the results from the MC Hamiltonian are in good agreement
with the analytical ones when $\beta > 1$.

Preliminary results for larger $N_{{\rm osc}}$ indicate that it is not necessary to increase $N$ accordingly. 
This property is  very important for a feasible application of the algorithm
to many body systems and QFT.

\section{Summary}

In this paper, we have extended the effective Hamiltonian method 
with a stochastic basis to QFT, and taken 
the Klein-Gordon model as an example.
The results are very encouraging.
We believe that the application 
of the algorithm to more complicated systems 
will be very interesting for non-perturbative investigation 
beyond the ground state.

\bigskip

\noindent
{\bf Acknowledgments}

We are grateful to the attendees of the workshop for useful
discussions.
X.Q.L. is supported by the
National Science Fund for Distinguished Young Scholars (19825117),
National Natural Science Foundation (10010310687),
Guangdong Provincial Natural Science Foundation (990212),
and the Ministry of Education of China.
H.K. has been supported by NSERC Canada.
We would also thank E.B. Gregory,
C.Q. Huang, J.Q. Jiang, J.J. Liu, and H. Xu for collaborations.

\end{document}